\documentclass[twocolumn,showpacs,preprintnumbers,amsmath,amssymb,superscriptaddress,floatfix]{revtex4}

\usepackage{graphicx}
\usepackage{dcolumn}
\usepackage{bm}

\begin{document}


\title{Experimental observation of the crystallization of a paired holon state}%

\author{A.~Rusydi}
\email{phyandri@nus.edu.sg}
\affiliation{Nanocore Institute, National University of Singapore, Singapore 117576}
\affiliation{Department of Physics, National University of Singapore, Singapore 117542}
\affiliation{Singapore Synchrotron Light Source, National University of Singapore, Singapore 117603}
\affiliation{Institut f\"{u}r Angewandte Physik, Universit\"{a}t Hamburg, Jungiusstrasse 11, 20355 Hamburg, Germany. Center for Free Electron Laser Science (CFEL), D-22607 Hamburg, Germany}

\author{W.~Ku}
\affiliation{Condensed Matter Physics \& Materials Science Department, Brookhaven National Laboratory, Upton, NY 11973}

\author{B.~Schulz}
\affiliation{Institut f\"{u}r Angewandte Physik, Universit\"{a}t Hamburg, Jungiusstrasse 11, 20355 Hamburg, Germany. Center for Free Electron Laser Science (CFEL), D-22607 Hamburg, Germany}

\author{R.~Rauer}
\affiliation{Institut f\"{u}r Angewandte Physik, Universit\"{a}t Hamburg, Jungiusstrasse 11, 20355 Hamburg, Germany. Center for Free Electron Laser Science (CFEL), D-22607 Hamburg, Germany}

\author{I.~Mahns}
\affiliation{Institut f\"{u}r Angewandte Physik, Universit\"{a}t Hamburg, Jungiusstrasse 11, 20355 Hamburg, Germany. Center for Free Electron Laser Science (CFEL), D-22607 Hamburg, Germany}

\author{D.~Qi}
\affiliation{Nanocore Institute, National University of Singapore, Singapore 117576}
\affiliation{Department of Physics, National University of Singapore, Singapore 117542}

\author{X.~Gao}
\affiliation{Department of Physics, National University of Singapore, Singapore 117542}

\author{A.T.S.~Wee}
\affiliation{Department of Physics, National University of Singapore, Singapore 117542}

\author{P.~Abbamonte}
\affiliation{Physics Department and Frederick Seitz Materials Research Laboratory, University of Illinois, Urbana, IL, 61801}

\author{H.~Eisaki}
\affiliation{Nanoelectronics Research Institute, AIST, 1-1-1 Central 2, Umezono, Tsukuba, Ibaraki, 305-8568, Japan}

\author{Y.~Fujimaki}
\affiliation{Department of Superconductivity, University of Tokyo, Bunkyo-ku, Tokyo 113, Japan}

\author{S.~Uchida}
\affiliation{Department of Superconductivity, University of Tokyo, Bunkyo-ku, Tokyo 113, Japan}

\author{M.~R\"{u}bhausen}
\affiliation{Institut f\"{u}r Angewandte Physik, Universit\"{a}t Hamburg, Jungiusstrasse 11, 20355 Hamburg, Germany. Center for Free Electron Laser Science (CFEL), D-22607 Hamburg, Germany}
\affiliation{Nanocore Institute, National University of Singapore, Singapore 117576}

\date{\today}

\begin{abstract}

A new excitation is observed at 201 meV in the doped-hole ladder cuprate
Sr$_{14}$Cu$_{24}$O$_{41}$, using ultraviolet resonance Raman scattering
with incident light at 3.7 eV polarized along the direction of the rungs.
The excitation is found to be of charge nature, with a temperature independent
excitation energy, and can be understood via an intra-ladder pair-breaking process.
The intensity tracks closely the order parameter of the charge density wave in the
ladder (CDW$_L$), but persists above the CDW$_L$ transition
temperature ($T_{CDW_L}$), indicating a strong local pairing above $T_{CDW_L}$.
The 201 meV excitation vanishes in La$_{6}$Ca$_{8}$Cu$_{24}$O$_{41+\delta}$,
and La$_{5}$Ca$_{9}$Cu$_{24}$O$_{41}$
which are samples with no holes in the ladders. Our results suggest that the doped holes
in the ladder are composite bosons consisting of paired holons that order below $T_{CDW}$.

\end{abstract}


\maketitle

\begin{figure}
\includegraphics[width=85.mm]{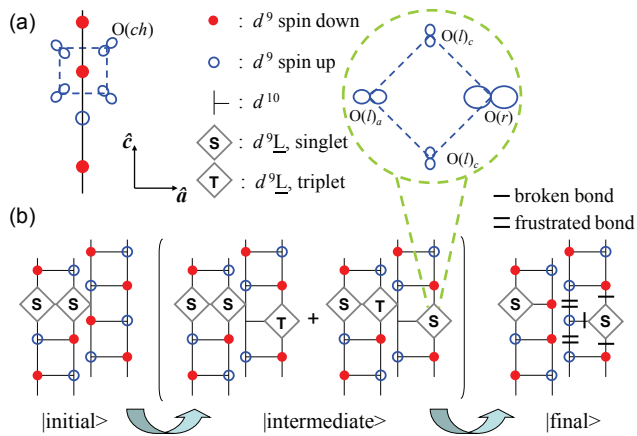}
\caption{(a) CuO$_2$ chain.
The charge transfer (CT),O($ch$) 2$p$$\rightarrow$Cu 3$d$ transition,
is polarization independent along the $a$- and $c$-axis\cite{Rusydi07}.
(b) Cu$_2$O$_3$ ladders consisting paired hole ``singlet" states and
Raman active processes, i.e. $|$intermediate$\rangle$, and $|$final$\rangle$ states.
(Zoom-in) Based on Wannier functions, the lowest energy state of holes in ladders is
a Zhang-Rice singlet and the weight of the orbital in the 4 oxygen atoms are different
due to the geometry of the ladder. Thus, the CT excitation is polarization dependent in $ac$ plane.
In the $|$final$\rangle$ state, the energy cost of each broken bond with a bar is J/4
and the each frustrated bond with a double-bar is J/2.
The two new bonds in the left ladder gives contributions cancel with each other.
}
\end{figure}

Important physics in strongly correlated materials is driven by the nature of the pairing
state of charges, i.e. holes in hole-doped materials such as  high-temperature
superconductor copper-oxides (cuprates) \cite{Kivelson03}. The paired hole states are believed to be
responsible for the mechanism of superconductivity (SC)
and the formation of the related but insulating charge density wave (CDW) order.
However, thus far the nature of pairing mechanism in both cases is still unclear.

An ideal system to study paired hole states is the two-leg ``spin ladder"
Sr$_{14-x}$Ca$_{x}$Cu$_{24}$O$_{41}$ (SCCO) which is believed to contain the basic physics of the
cuprates\cite{Dagotto92,Sigrist94,White02,Carr02,Nishimoto02,Wohlfeld07}. SCCO is a
self-doped material with 6 holes per formula unit and is a layered material
consisting of two different cuprates structures: CuO$_2$ ``chains" and Cu$_2$O$_3$
``ladders" (see Fig. 1 and Ref.\cite{McCarron88} for the complete structure).
We note, that x does not change the total number of holes, however it redistributes
the holes in the chain and ladder\cite{Rusydi07}.
SCCO has striking properties: it even exhibits superconductivity for
$x>10$ under pressure above 3 GPa\cite{Kojima02}.
For $x=0$, dc conductivity and low-frequency dielectric
measurements \cite{Blumberg02,Vuletic03} suggest the existence of
unconventional charge density waves (CDWs) which exhibit below $\sim$200 K
an energy gap of about $\sim$112 meV (=$\Delta_{CDW}$).

The distribution of holes in the chains and ladder has been one of the central subjects
as many interpretations depend on this distribution. For instance, neutron diffraction
results by Matsuda {\em et al.}\cite{Matsuda96} was thought to be a signature of a
superlattice reflection in the chain with periodicity $L_c= 0.25$ ($L_c$ is the
indices Miller of the chain). The neutron diffraction result was interpreted as a hole
modulation in the chain via a spin dimerization leading to 5 holes residing in the chains
and 1 hole in the ladders. However, more recent neutron diffraction studies by Braden
{\em et al.}\cite{Braden04}, analysis of the crystal structure by van Smaalen\cite{Smaalen03},
hard X-ray diffraction result by Zimmermann {\em et al.}\cite{Zimmermann06} and
resonant soft x-ray scattering (RSXS) studies \cite{Abbamonte04, Rusydi06, Rusydi08}
have clearly shown that the $L_c = 0.25$ peak is a structural modulation driven by a misfit
strain between the chains and ladders.

A direct way to measure the distribution of holes is x-ray absorption spectroscopy (XAS). However, this was also subject to interpretation
because the model used previously had unexplained discrepancies with regard to the strong
polarization dependence observed in XAS\cite{Nucker00}. A recent polarization dependent
XAS study on SCCO\cite{Rusydi07} has resolved these discrepancies and has accordingly
revisited the number of holes in the chain and ladders in which for $x$ = 0,
there 3.2 holes are in the chain and 2.8 in the ladder.
Furthermore, the combination of XAS and RSXS has revealed that:
(1) SCCO contains the unconventional CDWs, i.e.
a hole Wigner crystal (HC) in the ladder or CDW$_L$ ($x$=0,10,11, and 12)\cite{Abbamonte04, Rusydi06}, and a
4$k_F$-CDW in the chain or CDW$_c$($x$=0)\cite{Rusydi08},
and (2) suggesting the existence of paired hole states along the rung of the
ladders\cite{Rusydi07}. It is concluded that the interplay of lattice commensuration, Coulomb repulsion,
and geometric tiling of edge-shared ladders is responsible for the
CDW$_L$ and enforces a unique environment for the holes to pair (see Fig. 1).
In contrast, the misfit strain between the ladder and chain substructures is the driving force for
the chain 4$k_F$-CDW.

Several studies have found the intimate connection between the local physics of the quasi one-dimensional two-leg spin ladders and two-dimensional (2D) cuprates\cite{Devereaux07}. Studies of the two-magnon
(2M) excitation in SCCO using Raman scattering in visible have suggested that a nearest-neighbor exchange coupling
$J$ ($\sim$100 meV ) in the ladder is isotropic\cite{Gozar01}. A similar observation was found
in 2D cuprates \cite{Devereaux07}.
An inelastic neutron scattering study found a spin-liquid
state with a spin gap of 40 meV for $x=0$\cite{Matsuda96}.
This spin gap energy is also similar to the spin-resonance
mode seen in the 2D cuprates\cite{Devereaux07}.
However, it is unclear whether the spin gap is directly relevant
to the energy of the paired holes because neutron scattering is not directly sensitive to charge.
Recently, a pair density wave consisting of a hole pair was proposed to exist in
2D cuprates\cite{Chen02,Tesanovic04}. This has been followed by a recent evidence for paired charges
in underdoped cuprates using a scanning tunneling microscope\cite{Kohsaka07}.
The suggested pair-density wave scenario is in fact very similar to a quasi 1D system\cite{Dagotto92,Sigrist94}.
Thus, the understanding of the quasi 1D-two-leg spin ladder is of fundamental importance to understand the
physics of the cuprates. Here, we present a pair-breaking excitation of holes at 201 meV
($\sim$2$\Delta_{CDW_L}$), the intimate relationship between the paired hole states and the ladder-CDW, and the existence of preformed hole pairs by using ultraviolet resonance Raman scattering (UV-RRS) and RSXS.

\begin{figure}
\includegraphics[width=65.mm]{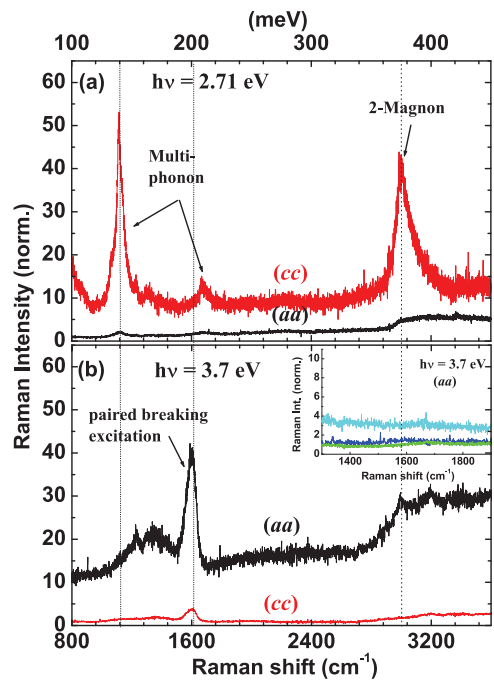}
\caption{Polarization dependence of Raman scattering (a) with
$h\nu$=2.7 eV, and (b) with $h\nu$=3.7 eV, for Sr$_{14}$Cu$_{24}$O$_{41}$.
The inset shows the Raman scattering for (blue) La$_{6}$Ca$_{8}$Cu$_{24}$O$_{41}$,
(green) La$_{6}$Ca$_{8}$Cu$_{24}$O$_{41.056}$, and (cyan) La$_{5}$Ca$_{9}$Cu$_{24}$O$_{41}$.
All data were taken at T = 20 K.
}
\end{figure}

Figure 2 shows inelastic light Raman scattering spectra with an incident photon
energy, $h\nu$, of (a) 2.7 and (b) 3.7 eV
for ($aa$) and ($cc$) polarizations of SCO and LCCO. (The ($xy$) polarization
geometry in Raman means that the incoming (out going) photons are polarized along $x$($y$)).
At $h\nu$=2.7 eV, our Raman result is similar to Ref.\cite{Gozar01}.
We have observed a two-phonon excitation at $\sim$138 meV and a three-phonon
excitation at $\sim$207 meV. These excitations are the second and
third order of a one-phonon excitation ($\sim$69 meV), respectively, that is originally from
Raman active Ag modes from oxygen and gets enhanced by Fr\"{o}hlich interaction\cite{Popovic00}.
They are strong in the ($cc$) polarization and weaker in the ($aa$).
We have also observed sharp and strong 2M excitation
in ($cc$), however weaker in ($aa$). The energy of the 2M excitation is nearly isotropic
at $\sim$375 meV Raman shift yielding an isotropic $J$ of $\sim$100 - 120 meV.

Our central observation is a new electronic excitation at 201 meV Raman
shift of SCO using UV-RRS measured with an incident photon energy of $h\nu$=3.7 eV (Fig. 2(b)).
The polarization dependence shows that the intensity of the 201 meV peak
in ($aa$) is at least 10 times higher than the intensity in ($cc$). In cross polarization,
the 201 meV feature is nearly invisible. Note, that the energy of this excitation is
about is twice the value of the CDW gap measured by dc conductivity and
low-frequency dielectric measurements \cite{Blumberg02,Vuletic03}.

Another striking evidence that the 201 meV is directly related to the presence of
holes comes from a doping dependent study. Reference samples
which contain no holes in the ladder but nearly identical magnetic structures as seen
by NMR and Neutron measurements,  do not show the 201 meV peak. \cite{Eccleston98,Kumagai97}.
If the 201 meV feature would be related to the spin, then it should also appear in the reference
samples. However, we have found that the 201 meV peak vanishes in L6C8CO$_{41}$,
L6C8CO$_{41.056}$, and L5C9CO, i.e. samples without holes in the ladders (see Fig. 2(b)).

\begin{figure}
\includegraphics[width=65.mm]{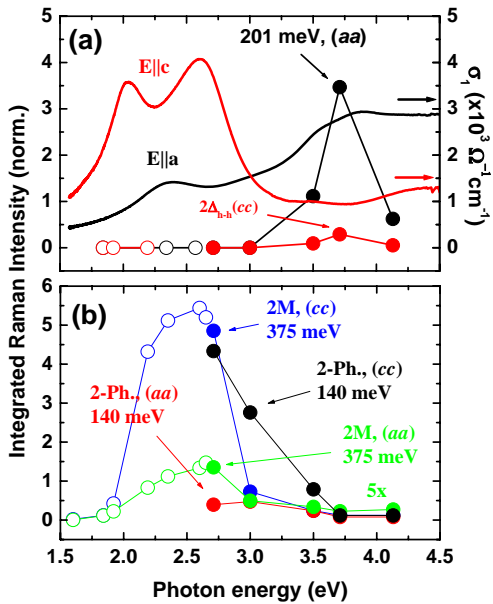}
\caption{(a) Resonance profile (RP) of the 201 meV excitation for ($aa$) and ($cc$) polarizations and
comparison with the optical conductivity, $\sigma_1$, for $a$ and $c$ polarizations.
Open-black and open-red circles are for ($aa$) and ($cc$) polarizations
taken from Ref.\cite{Gozar01}, respectively. (b) RP of the two-magnon (375 meV)
and the multiple phonon excitation (140 meV) shows that these features are weaker at higher
$h\nu$. Open-blue-cirlces and open-green-cirlces are ($aa$) and ($cc$) polarizations
of 2M from Ref.\cite{Gozar01}, respectively, however corrected for the approriate
dielectric function\cite{footnote1}. The data all are taken at 20 K.
}
\end{figure}

Furthermore, we have studied a complete resonance profile (RP) of
201 meV, 375 meV, and 140 meV excitations which are shown in Figure 3(a)$\&$(b).
The RP of these excitations is shown in Fig. 3(a)$\&$(b) which identifies the involved Raman matrix elements.
The 201 meV resonates at $\sim$3.7 eV. This is in contrast to the RP
of the 2M which is in resonance at $\sim$2.5 eV and is off-resonance at $\sim$3.7 eV
for both polarizations. This shows that the 2M and the 201 meV excitation have different
matrix elements. Furthermore, a novel unconventional magnetic excitation would require, due to spin conservation,
a large spin/orbit coupling, which is small for the  Cu3$d^9$
ion\cite{Gozar04}. On the other hand, the RP of the multi phonon
excitation is also very different compared to the RP of 201 meV excitation.
The intensity of 140 meV is  getting weaker with increasing incident photon energy.
Thus, the 201 meV feature is neither of magnetic nor of phonon origin.

The matrix element of the 201 meV excitation with a resonance at 3.7 eV can be well-explained in a two-step pair-breaking process illustrated in Fig. 1(b).  First, in the ladder next
to an existing paired holes of singlet state, $d^9\b{L}(S)d^9\b{L}(S)$, the $a$-polarized 3.7 eV
photon creates a charge transfer, $d^{10}d^9\b{L}(T)$, along the $a$-direction, leading to an
virtual intermediate state.  (Here ($T$) indicates triplet state, as 3.7 eV is typical energy for
creation of a triplet state).  The intermediate state is a many-body state consisting of multiple
determinants, including $d^9\b{L}(S)d^9\b{L}(S) + d^{10}d^9\b{L}(T)$,
$d^9\b{L}(S)d^9\b{L}(T) + d^{10}d^9\b{L}(S)$, and other permutation of the holes that are
strongly mixed due to their nearly identical energy.  Second, the intermediate state releases
the energy via an inter-ladder charge transfer, leaving one singlet on each ladder.  The Raman
shift of this pair-breaking process is ~$7J/4$ (ignoring Coulomb repulsion and inter-chain
magnetic coupling), in reasonable agreement with current lore of the strength of intra-ladder
magnetic coupling $J$ (c.f.: 2(b)).
Not surprisingly, such pair-breaking excitation is absent in undoped
LCCO where no paired holes reside and should be $a$-polarized in the Raman experiment.

The observed resonant Raman process has strong implications to the electronic structure of
the CDW$_L$ phase and more generally of the doped holes in the ladder. Specifically, it indicates that
the CDW$_L$ phase consists of paired holes residing across the rungs. Furthermore, across the
paired holes the antiferromagnetic spin configurations are anti-phased, i.e. $\pi$-phase shifted,
similar to those in the ``stripe" phase of perovskite cuprates\cite{Tranquada95}.
It is straightforward to verify that initial states without this phase shift would have different excitation
energies by at least $J$. In the ladder, this $\pi$-phase shift
is apparently driven by the kinetic energy of the holes, similar to the holon propagation in
pure 1D system, as only with the $\pi$-phase shift the paired holes can move freely along
the ladder without causing magnetic frustration.  In essence, the paired holes in the ladder
can be regarded as a composite boson consisting of a pair of holons at low temperature.

Further insights can be obtained from the temperature dependence of the pair-breaking excitation and
its contribution to the formation of the CDW$_L$. Figure 4(a) shows the Raman
scattering spectra for selected temperatures.  The reduction of the intensity at higher temperatures
is apparent, as well as the constant energy of the pair-breaking excitation.  This indicates strongly
that the energy scale of the local pairing is much larger than that of the CDW$_L$ ordering, and is not
temperature dependent.  Interestingly, when comparing (c.f.: Fig. 4(b)) the intensity of the
pair-breaking excitation with the CDW$_L$ order parameter, as measured by the integrated intensity of the
RSXS CDW Bragg peak,  both track each other very closely below $T_{CDW_L}$.
This observation confirms the intimate dependence to the phase coherence of the CDW$_L$
order parameter and the $\pi$-phase shifted antiferromagnetic background. Furthermore, above
$T_{CDW_L}$ where the CDW$_L$ order vanishes,  but the intensity of the pair-breaking excitation remains finite and almost temperature independent. In consequence this has to be seen as evidence for the existence of a disordered paired hole states above
$T_{CDW_L}$, and the short-range antiferromagnetic order with a $\pi$-phase shift (c.f.: 4(c)).

\begin{figure}
\includegraphics[width=85.mm]{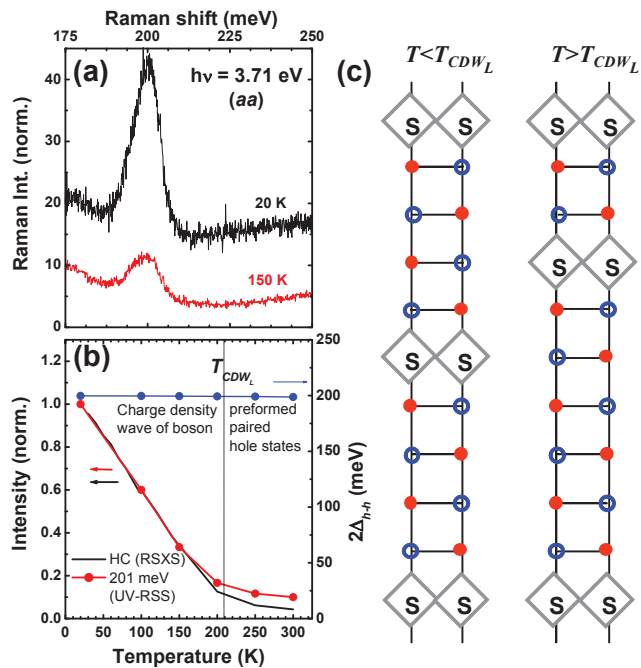}
\caption{(a) Raman scattering for selected temperatures, 20 K and 150 K.
(b) Comparison of the intensity of the 201 meV excitation and the Bragg peak of the CDW$_L$
measured with RSXS as function of temperature.
(c) A cartoon shows the intimate connection between CDW$_L$ and paired hole states.
At $T>T_{CDW_L}$ preformed paired holes already exist, however their phases are incoherent.
Upon cooling, $T<T_{CDW_L}$, the paired hole states are crystallized and develop
a CDW$_L$ long range order. The periodicity of the CDW$_L$ is 5 ladder unit ($c_L$) as seen with
RSXS\cite{Abbamonte04,Rusydi08}.
}
\end{figure}

In this context it is important to outline the connection between the energy scales of
the 201 meV excitation, the dc conductivity, the low-frequency dielectric measurements\cite{Blumberg02,Vuletic03},
and angle-resolved photoemission in 2D cuprates\cite{Devereaux07}.
Firstly, the dc conductivity and the low-frequency dielectric measurements
show that the activation energy of SCO is $\sim$112 meV ($\propto \Delta_{CDW}$), which half of the energy of the pair-breaking excitation($\propto 2 \Delta_{CDW}$).
In some theories the energy of the paired hole state would be identical to the spin-gap
energy\cite{Roux05}. However, here we find that the $local$ pairing energy of the holes
is 5 to 6 times higher than the spin gap.
Moreover, it is also remarkable that the pair-breaking energy of 201 meV is about twice of
pseudogap in 2D cuprates as observed with ARPES\cite{Devereaux07}.
This similarity in energy scales supports the notion that the pseudo-gap reflects the energy
scale of a local pair in the absence of phase-coherence.
Furthermore, the persistence of large pair-breaking energy above $T_{CDW}$ in our measurement
supports the existence of pre-formed pairs at temperature much larger than $T_{CDW}$.

In conclusion, we have observed the pair-breaking excitation of holes at 201 meV Raman shift
along the rung of the ladder of SCO. Below $T_{CDW_L}$, the paired hole states are crystallized and
responsible for formation of the CDW$_L$long range order in the ladder. Above $T_{CDW_L}$, the paired hole
states still exist as preformed states, however they are disordered.
The CDW$_L$ and paired holes states occur in a unique environment of $\pi$-phase shifted
antiferromagnetic spins similar to stripe phase in 2D cuprates. Our results open a new possibility
to study paired hole states in systems close to the instability
towards the formation of the CDW$_L$ and to understand competing order parameters in correlated
materials such as in the high temperature superconductors.

We would like to acknowledge intense discussions with M.V. Klein, G.A. Sawatzky, and S.L. Cooper.
This work was supported by VH-FZ-007, DFG Ru 773/3-2,
NRF2008NRF-CRP002-024, NUS
YIA, NUS cross faculty, FRC, NUS Advanced Functional Materials (R-263-000-432646),
the 21st Century COE program of the Japan Society for Promotion of Science, U.S. DOE
Grant No. DE-FG02-06ER46285 and theoretical support DOE-CMSN under Contract No. DE-AC02 98CH10886.
Work partly performed at SSLS under NUS Core Support C-380-003-003-001, A*STAR/MOE RP
3979908M and A*STAR 12 105 0038 grants.


\begin{references}
\bibitem{footnote1} In the previous studies, all Raman spectra were corrected for the room temperature dielectric function only, which was assumed to be nearly temperature independent.
\end{references}
\end{document}